\documentclass{aa}  

\DeclareRobustCommand{\VAN}[3]{#2}
\let\VANthebibliography\thebibliography
\def\thebibliography{\DeclareRobustCommand{\VAN}[3]{##3}\VANthebibliography}

\usepackage{caption}
\usepackage{subcaption}
\usepackage{mhchem}
\usepackage{graphicx}	

\usepackage{amsmath}	
\usepackage{amssymb}	
\usepackage{array}
\usepackage{multirow}
\usepackage{bigdelim}
\usepackage[varg]{txfonts}

\graphicspath{{figures/}}

\defcitealias{Padovani2022}{P22}
\defcitealias{Bialy2020}{B20}

\begin{document}

\title{Cosmic-ray-induced \ce{H2} line emission}
\subtitle{Astrochemical modeling and implications for JWST observations}
\titlerunning{Cosmic-ray induced \ce{H2} emission}
\authorrunning{Gaches et al.}

\author{Brandt A. L. Gaches
        \inst{1}\fnmsep\inst{2}\thanks{E-mail: gaches@ph1.uni-koeln.de}
        \and 
        Shmuel Bialy
        \inst{3}
        \and
        Thomas G. Bisbas
        \inst{1}\fnmsep\inst{4}
        \and
        Marco Padovani
        \inst{5}
        \and
        Daniel Seifried
        \inst{1}
        \and
        Stefanie Walch
        \inst{1}}
\institute{
I. Physikalisches Institut, Universit\"{a}t zu K\"{o}ln, Z\"{u}lpicher Stra{\ss}e 77, 50937, K\"{o}ln, Germany
\and
Center of Planetary Systems Habitability, The University of Texas at Austin,  USA
\and
Department of Astronomy, University of Maryland, College Park, MD 20742, USA
\and
Department of Physics, Aristotle University of Thessaloniki, GR-54124 Thessaloniki, Greece
\and
INAF–Osservatorio Astrofisico di Arcetri, Largo E. Fermi 5, 50125 Firenze, Italy
}

\date{Accepted XXX. Received YYY; in original form ZZZ}

\abstract
{It has been proposed that \ce{H2} near-infrared lines may be excited by cosmic rays and allow for a determination of the cosmic-ray ionization rate in dense gas. One-dimensional models show that measuring both the \ce{H2} gas column density and \ce{H2} line intensity enables a constraint on the cosmic-ray ionization rate as well as the spectral slope of low-energy cosmic-ray protons in the interstellar medium (ISM).}
{We aim to investigate the impact of certain assumptions regarding the \ce{H2} chemical models and ISM density distributions on the emission of cosmic-ray induced \ce{H2} emission lines. This is of particular importance for utilizing observations of these lines with the James Webb Space Telescope to constrain the cosmic-ray ionization rate. }
{We compare the predicted emission from cosmic-ray induced, rovibrationally excited \ce{H2} emission lines for different one- and three-dimensional models with varying assumptions on the gas chemistry and density distribution.}
{We find that the model predictions of the \ce{H2} line intensities for the (1-0)S(0), (1-0)Q(2), (1-0)O(2) and (1-0)O(4) transitions at 2.22, 2.41, 2.63 and 3.00 $\mu$m, respectively, are relatively independent of the astro-chemical model and the gas density distribution when compared against the \ce{H2} column density, making them robust tracer of the cosmic-ray ionization rate.}
{We recommend the use of rovibrational \ce{H2} line emission in combination with estimation of the cloud's \ce{H2} column density, to constrain the ionization rate and the spectrum of low energy cosmic-rays.
}


\maketitle

\section{Introduction}
Low-energy cosmic rays (LECRs) with energies less than 1 GeV play a significant role in driving the thermochemistry of the molecular interstellar medium (ISM) \citep{dalgarno2006, indriolo2013}. In regions shielded from ultraviolet (UV) radiation, LECRs are the dominant source of ionization. The ionization they provide drives a rich ion-neutral chemistry, leading to the formation of many astronomically important molecules, as well as the initiation of deuteration \citep{bayet2011, caselli2012, indriolo2013, bialy2015, grenier2015}. Further, LECRs provide an important source of heating, and through the ionization fraction regulate the impact of non-ideal magnetohydrodynamic effects such as ambipolar diffusion \citep{padovani2020}. 

Determining the flux of LECRs irradiating molecular clouds is a difficult endeavor. There have been a number of investigations using a range of molecular lines \citep[e.g.][]{Caselli1998, vanderTak2000, McCall2002, mccall2003, hezareh2008, shaw2008, ceccarelli2011,hollenbach2012, Indriolo2012, Ceccarelli2014, Podio2014, vaupre2014, cleeves2015, Indriolo2015, lepetit2016, Fontani2017, Neufeld2017, Favre2018, indriolo2018, bacalla2019, Barger2020, bovino2020, redaelli2021} and gas temperature \citep[e.g.][]{ivlev2019} to estimate the cosmic-ray ionization rate (CRIR), denoted as $\zeta$. In diffuse gas, absorption studies of simple molecular ions probe the CRIR. However, dense gas measurements typically rely on astrochemical modeling and thus are prone to a number of degeneracies, in particular the treatment of the CRIR \citep{gaches2019}.

Recently, \citet[][hereafter B20]{Bialy2020} proposed a novel method to estimate the LECR flux using H$_2$ rovibrational line emission. As the primary CR protons penetrate into the cloud they produce a population of secondary electrons which efficiently excite the rovibrational transitions of \ce{H2} (especially of the first vibrational state $v=1$) resulting in H$_2$ line emission in the near-IR. As shown by \citet{Bialy2022} and \citet[][hereafter P22]{Padovani2022}, observations of H$_2$ rovibrational lines may be used to constrain the spectrum of LECRs that is prevailing in the ISM.

The \ce{H2} lines of interest are shown in Table \ref{tab:h2params}, between 2.22 and 3 $\mu$m. The James Webb Space Telescope (JWST) will be able to observe these lines with the NIRSPEC instrument simultaneously. The unprecedented observations will enable JWST to determine the CRIR in dense molecular gas where  absorption measurements are difficult. As such, exploring how different model assumptions impact the line predictions is crucial. 

The aforementioned previous calculations assumed a fully molecular one-dimensional slab which enabled parameter-space predictions of the \ce{H2} line intensity as a function of the observed \ce{H2} column density, $N_{\rm obs}$. These calculations assumed fully molecular clouds and did not include the effects of FUV photodissociation and an inhomogeneous density structure, which result in regions in the cloud that are partially atomic. In addition, as previous models are one-dimensional, they assume that the observed column density along the LOS and the effective column density that attenuates CRs as they penetrate into the cloud, are identical. In an inhomogeneous three-dimensional  cloud, CRs can penetrate from different directions, along ``rays" passing through different density profiles (not only along the direction of the LOS), resulting in strong fluctuations in the local CR ionization and excitation rate. Therefore, the role of density structure and chemical evolution model (e.g. equilibrium versus non-equilibrium), should be constrained, as these will impact the conversion of the local quantity (induced \ce{H2} emission) to an integrated quantity (observed \ce{H2} line intensity).

In this paper, we present synthetic \ce{H2} line emission  (2D plane-of-the-sky) maps of a realistic molecular cloud irradiated by an interstellar CR proton spectrum. We use the three-dimensional astrochemical models presented in \citet{Gaches2022}, which include a prescription for CR attenuation and self-consistently formed molecular clouds from the SILCC-Zoom project \citep{Seifried2017}, and the CR excitation rates computed in \citetalias{Padovani2022}.

\begin{table}[ht!]
    \caption{\ce{H2} transitions and physical constants}
    \label{tab:h2params}
    \centering
    \begin{tabular}{c|ccccc}
        Transition & $J_u$ & $J_l$ & $\lambda$ ($\mu$m) & $E_{ul}$ (eV) & $\alpha_{ul}$ \\
        \hline
        (1-0)S(0) & 2 & 0 & 2.22 & 0.56 & 0.30 \\
        (1-0)Q(2) & 2 & 2 & 2.41 & 0.51 & 0.36 \\
        (1-0)O(2) & 0 & 2 & 2.63 & 0.47 & 1.00 \\
        (1-0)O(4) & 2 & 4 & 3.00 & 0.41 & 0.34 
    \end{tabular}
\end{table}

\section{Methods}\label{sec:methods}
\begin{table*}[ht!]
    \renewcommand{\arraystretch}{1.25}
    \caption{Physical and chemical models}
    \label{tab:models}
    \centering
    \begin{tabular}{l@{\,}c||>{\centering}m{7em}|>{\centering}m{3em}|>{\centering}m{5.5em}|c|>{\centering}m{12em}}
        & Model &  Density & FUV & CR Attenuation & Code &  Notes \tabularnewline
        \cline{2-7}
        \ldelim\{{5}{*}[{\rotatebox[origin=c]{90}{1D}}] & 1 & Constant \\ $n$ = 10 cm$^{-3}$ & 1 & $\checkmark$ & {\sc 3D-PDR}&  \tabularnewline
        \cline{2-7}
        & 2 & Constant \\ $n$ = 10$^3$ cm$^{-3}$ & 1 & $\checkmark$ & {\sc 3D-PDR} &  \tabularnewline
        \cline{2-7}
        & 3 & Variable\\ $N_{\rm eff}$ -- $n$  & 1 & $\checkmark$ & {\sc 3D-PDR} & Following Eq.~(\ref{eq:avn})  \tabularnewline
        \cline{2-7}
        \ldelim\{{3}{*}[{\rotatebox[origin=c]{90}{3D}}]& 4 & Variable \\ Simulation  & 10 & $\checkmark$ & {\sc 3D-PDR} &  \citet[]{Wu17, bisbas2021}\tabularnewline
        \cline{2-7}
        & 5 & Variable \\ Simulation  & 1.4 & &{\sc Flash} &  SILCC-Zoom, \citet[]{Seifried2017}\tabularnewline
    \end{tabular}
\end{table*}
We model a molecular cloud that is impacted by a flux of cosmic rays, and calculate the resulting H$_2$ rovibrational excitation and the consequent NIR line emission from the cloud.

\subsection{Incident CR flux}
For the CRs that are impinging on the cloud surface we assume the interstellar CR proton spectrum from \citet{padovani2018}, with a 
low-energy spectral slope of $\alpha=-0.8$; this is the ``$\mathcal{H}$ model'' which provides a good agreement to observations of the CRIR in diffuse clouds.

\subsection{CR attenuation}
As the CRs penetrate into the cloud they lose energy through ionization, dissociation and excitation. We account for this attenuation process by adopting the depth-dependent CRIR, $\zeta(N_{\rm eff})$, from Table F.1 of \citet{padovani2018}, as well as a depth-dependent H$_2$ excitation rate, $\zeta_{\rm ex}(N_{\rm eff})$ from \citetalias{Padovani2022} (see their Figs. 5, 6 and also Fig. 3 in \citealt{Bialy2022}). Hereafter we consider a set of cloud models, including 1D slab geometry models, as well as 3D models based on hydro simulations of turbulent clouds, as summarized in Table \ref{tab:models}. For our 1D models, 
$N_{\rm eff}(z)= \mu N'(z)$
where $N'(z) = \int_0^z n dz'$ is the column density from cloud edge to the point of interest at depth $z$ inside the cloud, and $\mu=\cos(\theta)$ is the cosine angle of the $B$-field lines with the cloud normal. We adopt $\mu=1$. In our 3D models, we utilize the effective column density by accumulating the column along {\sc Healpix} rays \citep{gorski2005},
\begin{equation}
    N_{\rm eff}(x,y,z) = -\frac{1}{2.5} \ln \left ( \frac{1}{N_{\ell}} \sum_{i=1}^{N_{\ell}} {\rm e}^{-2.5 N_{i}(x,y,z)'} \right ),
\end{equation}
for $N_{\ell} = 12$ rays at the {\sc Healpix} $\ell = 0$ level of refinement. 

\subsection{Density structure}
To explore the effect of the cloud structure on the resulting H$_2$ line emission, we consider five models with different density distributions and chemical properties, as summarized in Table \ref{tab:models}. 

Models 1 -- 3 are one-dimensional slabs. Models 1 and 2 assume a constant density, of $n=10$ and 10$^3$ cm$^{-3}$, respectively. Model 3 has a variable density profile in which the density and column density at each point in the cloud are related through 
\begin{equation}
\label{eq:avn}
    N_{\rm eff}(z) = 8.05 \times 10^{19} \exp \left [1.6 \left ( \frac{n}{{\rm cm}^{-3}} \right )^{0.12} \right ] \ {\rm cm^{-2}} . 
\end{equation}
This relies on the empirical $A_{\rm V} - n$ relation found by  \citet{bisbas2019} and Bisbas et al. (in prep) based on a series of turbulent ISM box simulations and galaxy disk simulations. For the densities significant for our results, this relationship well reproduces the average densities as a function of effective column density (see Fig. 5 of \citet{Gaches2022}). This relationship is for a Solar metallicity gas, and can change with metallicity \citep{hu2021}. Appendix \ref{app:oneD} provides additional details on the 1D astrochemical models. 

We also use two three-dimensional density distributions (Models 4 and 5). Model 4 uses the density distribution and astrochemical model from \citet{Gaches2022}, which was also previously used in \citet{bisbas2021}. This cloud (called ``dense'' cloud) is a subregion from the larger-scale simulations of \citet{Wu17}. The ``dense'' cloud is located in a cube with uniform resolution of 112$^3$ cells, a side length $L = 13.88$ pc, total mass $M_{\rm tot} = 5.9\times10^4$ M$_{\sun}$, and mean H-nucleus density $\left< n \right> = 640$ cm$^{-3}$ (see \citealt{Wu17} and \citealt{bisbas2021} for more details). Model 5 is a molecular cloud from the magneto-hydrodynamic (MHD) SILCC-Zoom simulations \citep{Seifried2017, Seifried2020}. These simulations model zoomed-in regions of the stratified disk SILCC simulations \citep{walch2015, girichidis2016} with the initial Galactic-scale magnetic field set to 3 $\mu$G and uses the {\sc Flash} 4.3 MHD code \citep{fryxell2000}. The SILCC-Zoom MHD cloud is located in a cube with side length,  $L\approx$ 125 pc with a total mass, $M_{\rm tot} = 2.13\times10^{5}$ M$_{\odot}$.

\subsection{Astrochemical Models}
For Models 1 -- 4, the chemistry is computed with a modified version of the public astrochemistry code {\sc 3d-pdr} \citep{bisbas2012, Gaches2022}. We use a reduced network derived from the UMIST 2012\footnote{\url{http://udfa.ajmarkwick.net}} chemical network database \citep{McElroy13} consisting of 33 species and 330 reactions. The chemistry is then evolved to steady-state using an integration time of 10~Myr. Models 1 -- 3 use an external FUV radiation field of $G_0 = 1$ (normalized to the spectral shape of \citealt{Habing68}) to minimize the impact of photochemistry and Model 4 uses $G_0 = 10$ to be consistent with previous studies \citep{bisbas2021, Gaches2022}.

Model 5 uses non-equilibrium chemistry with a network of 7 species \citep{glover2007a, glover2007b, glover2010, glover2012}, a constant atomic hydrogen CRIR $\zeta = 3\times10^{-17}$ s$^{-1}$ and an external FUV radiation field $G_0 = 1.4$. The effective column density is computed and stored during the simulation as described above using the {\sc TreeRay/OpticalDepth} module \citep{clark2012, walch2015, wunsch2018}. Due to the use of a constant CRIR, the chemistry is not entirely self-consistent with our treatment of the excitation rate, as described below. However, the ionization rates we consider are not high enough to greatly impact the \ce{H2} abundances. Therefore, our main results will not be significantly altered by this assumption.

\subsection{\ce{H2} excitation and line emission}

In steady state, the flux of secondary electrons becomes independent of the local density \citep[see][]{Ivlev2021}. Thus, we can use the calculation of the excitation rate $\zeta_{\rm exc,u}$ from \citetalias{Padovani2022} for $v_u = 1$, $J_u = 0, 2$. For the ``$\mathcal{H}$'' cosmic-ray flux model, the excitation rate varies from 10$^{-15}$ to 10$^{-16}$ s$^{-1}$ between the cloud surface and interior \citepalias[see Fig. 5 in][]{Padovani2022}. This calculation uses the CR energy loss function assuming a fully molecular gas. In practice, the loss function should account for a mix of atomic and molecular hydrogen, however as we show in the appendix, this has a marginal impact on our results for $N_{\rm obs} > 10^{21}$ cm$^{-2}$ (see Figure \ref{fig:1D_h2_abunds}).

Given, $\zeta_{\rm exc,u}$, the emissivity for a specific \ce{H2} line is
\begin{equation}
    \epsilon_{ul} = \alpha_{ul} \frac{E_{ul}}{4\pi} \zeta_{\rm exc,u} n(\ce{H2})  \,,
\end{equation}
where $\alpha_{ul}$ is the probability that the excitation of state $(v_u, J_u)$ will be followed by radiative decay to state $(v_l, J_l)$, $E_{ul}$ is the transition energy, and $n(\ce{H2})$ is the \ce{H2} number density. The $\alpha_{ul}$ factor does not include collisional quenching as the densities we consider lie below the critical density (e.g., $n_{\rm crit} \approx 10^{11}$ cm$^{-3}$ at 100 K; \citealt{Bialy2020}). Our models assume \ce{H2} is entirely in the para- state, which is applicable for the dense regions were are primarily concerned with (\citealt{flower2006}, see Appendix \ref{app:oneD} for an exploration of the impact of a different ortho-to-para ratio).

The line-of-sight integrated line intensity is then:
\begin{equation}
    F_{ul}(x,y) = \int_0^{L} \epsilon_{ul}(x,y,z) e^{-\sigma_{d} N'(x,y,z)} {\rm d}z \,,
\end{equation}
where $N'(x,y,z) = \int_0^z n(x,y,z') {\rm d}z'$ is the cumulative column density along the line of sight, from the cloud edge to a point inside the cloud (at depth $z$), $L$ is the cloud size, and
$\sigma_d = 4.5\times10^{-23}$ cm$^2$ is the NIR dust absorption cross section per hydrogen nuclei (\citetalias{Bialy2020}, \citealt{draine2003}).
Hereafter, we also define the {\it total} column density integrated along the LOS, $N_{\rm obs} \equiv \int_0^L n(x,y,z') {\rm d}z'$. Similarly, $N_{\rm obs}(\ce{H2})$ is the LOS-integrated column density of H$_2$.

\section{Results}\label{sec:results}
\begin{figure*}[h!]
    \centering
    \includegraphics[width=0.95\textwidth]{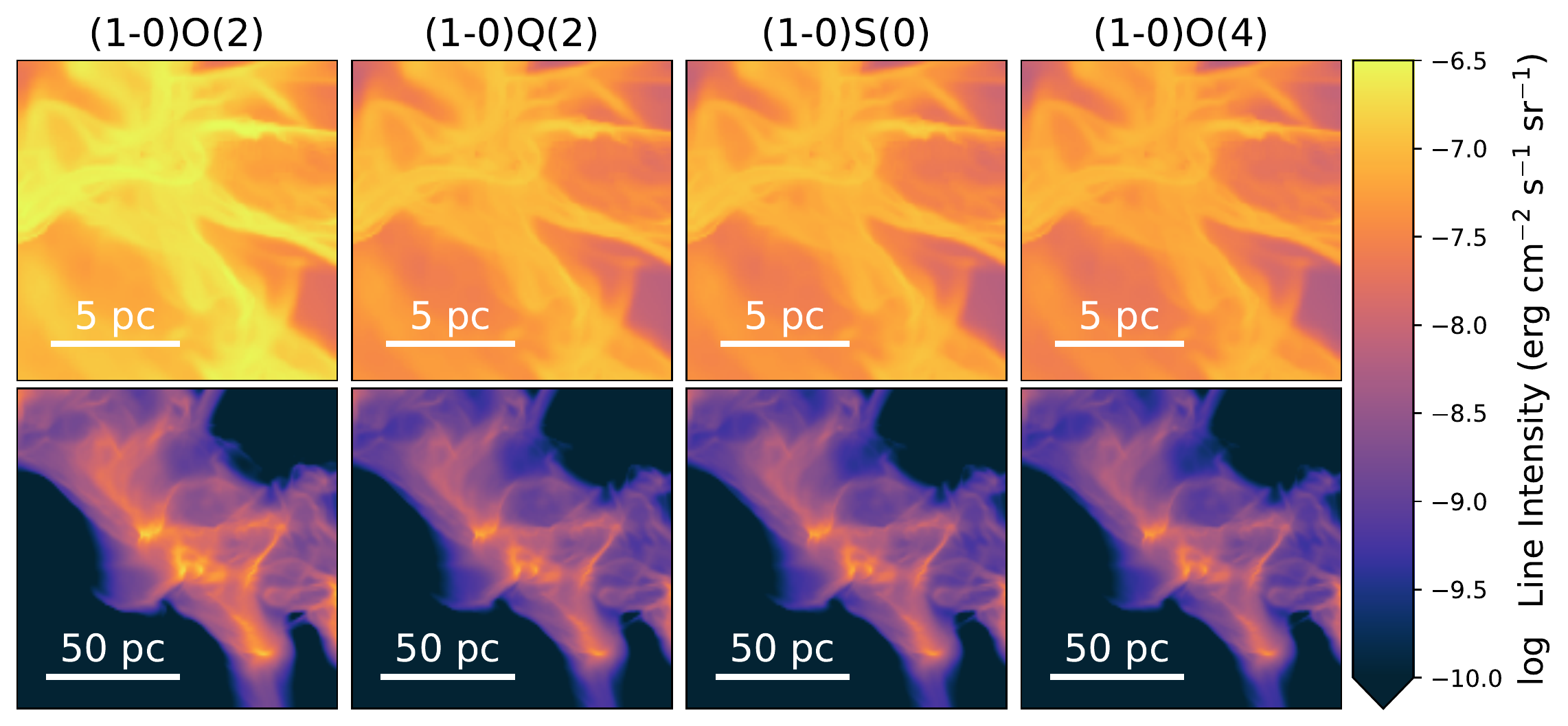}
    \caption{\label{fig:flxMaps}Line-of-sight line intensities of the {\sc 3d-pdr} model (top; Model 4) and SILCC-Zoom MHD cloud (bottom; Model 5) for the four different \ce{H2} lines in Table \ref{tab:h2params}.}
\end{figure*}

Figure \ref{fig:flxMaps} shows the line intensity of the denoted \ce{H2} line, seen along the $z$-axis, for Models 4 (top) and 5 (bottom). The observed fluctuations in the line intensities correspond to density fluctuations in the cloud, as well as variations in the effective column density. The emission saturates at high column densities due to the obscuration of dust. We note that these clouds formed through different processes: the Model 4 cloud is the product of a cloud-cloud collision and the Model 5 cloud is likely the result of supernova shells interacting  and contains structures on larger length scales, and thus also more diffuse gas. 

\begin{figure}
    \centering
    \includegraphics[width=0.5\textwidth]{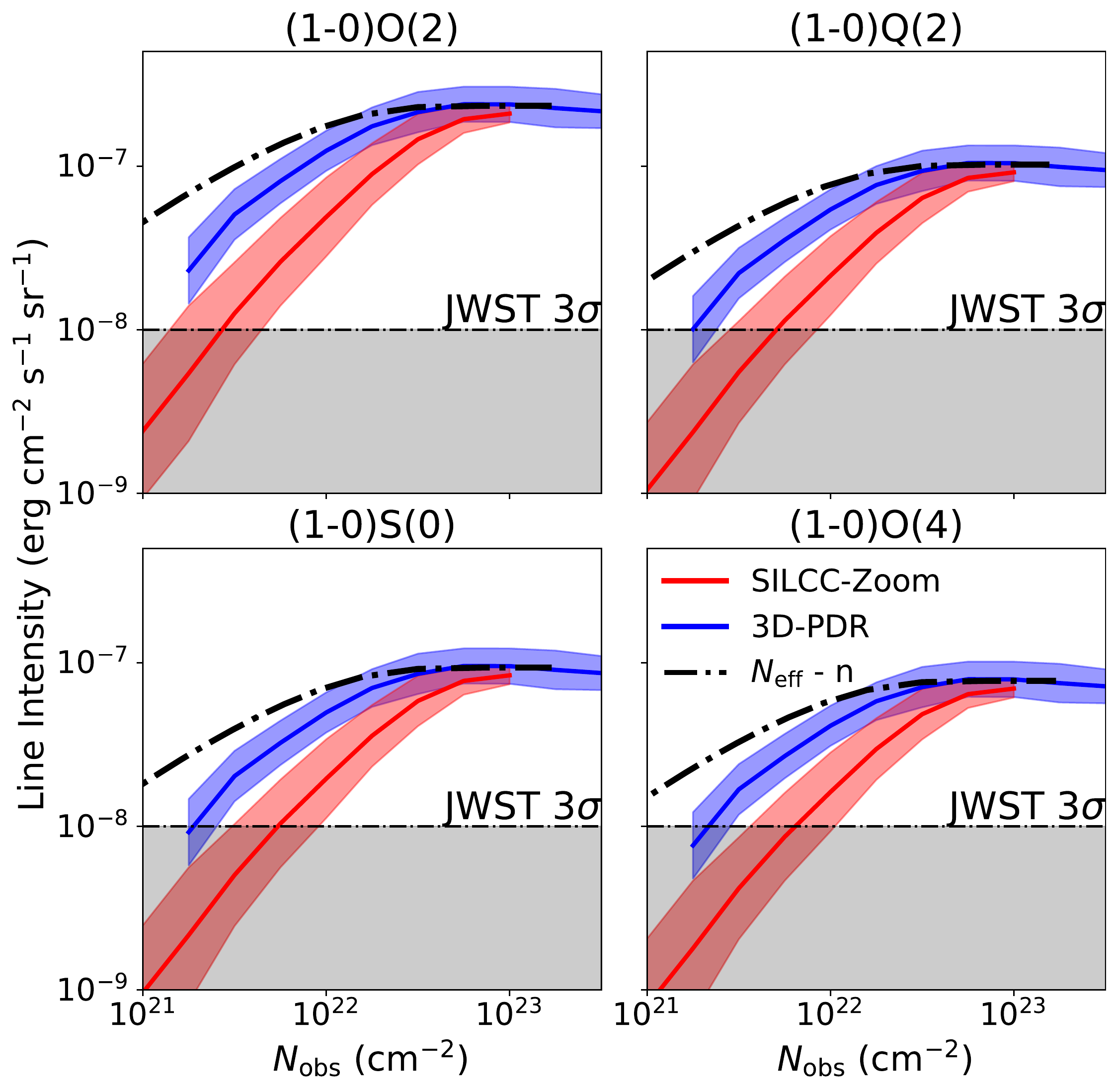}
    \caption{\label{fig:flxDistHMHD}Logarithmic column-density bin-averaged line intensities versus the total H-nucleus column density, $N_{\rm obs}$ for the \ce{H2} lines in Table \ref{tab:h2params}. The blue and red lines correspond to Models 4 and 5, respectively with the filled regions representing $\pm 2\sigma$. Black dashed-dot line shows the results for model 3. The shadowed region shows the JWST sensitivity with a signal-to-noise ratio of 3 with 1.25 h of integration and 50 shutters \citepalias{Padovani2022}.}
\end{figure}

Figure \ref{fig:flxDistHMHD} shows the line intensities as a function of the total integrated column of hydrogen nuclei, $N_{\rm obs}$, for our three-dimensional models (models 4,5) and the $N_{\rm eff}$ -- $n$ 1D model (model 3). We find that these models rapidly diverge for $N_{\rm obs} < 10^{22}$ cm$^{-2}$. The divergence is caused by the substantial differences in H-H$_2$ chemical structure. In particular, Model 5 is more diffuse than Model 4, and while both Models 3 and 5 evolve the chemistry to steady state, Model 4 evolves the chemistry with a non-equilibrium solver. As such, the models exhibit different \ce{H2} abundance distributions, driving the divergence at low column densities. 

\begin{figure}
    \centering
    \includegraphics[width=0.5\textwidth]{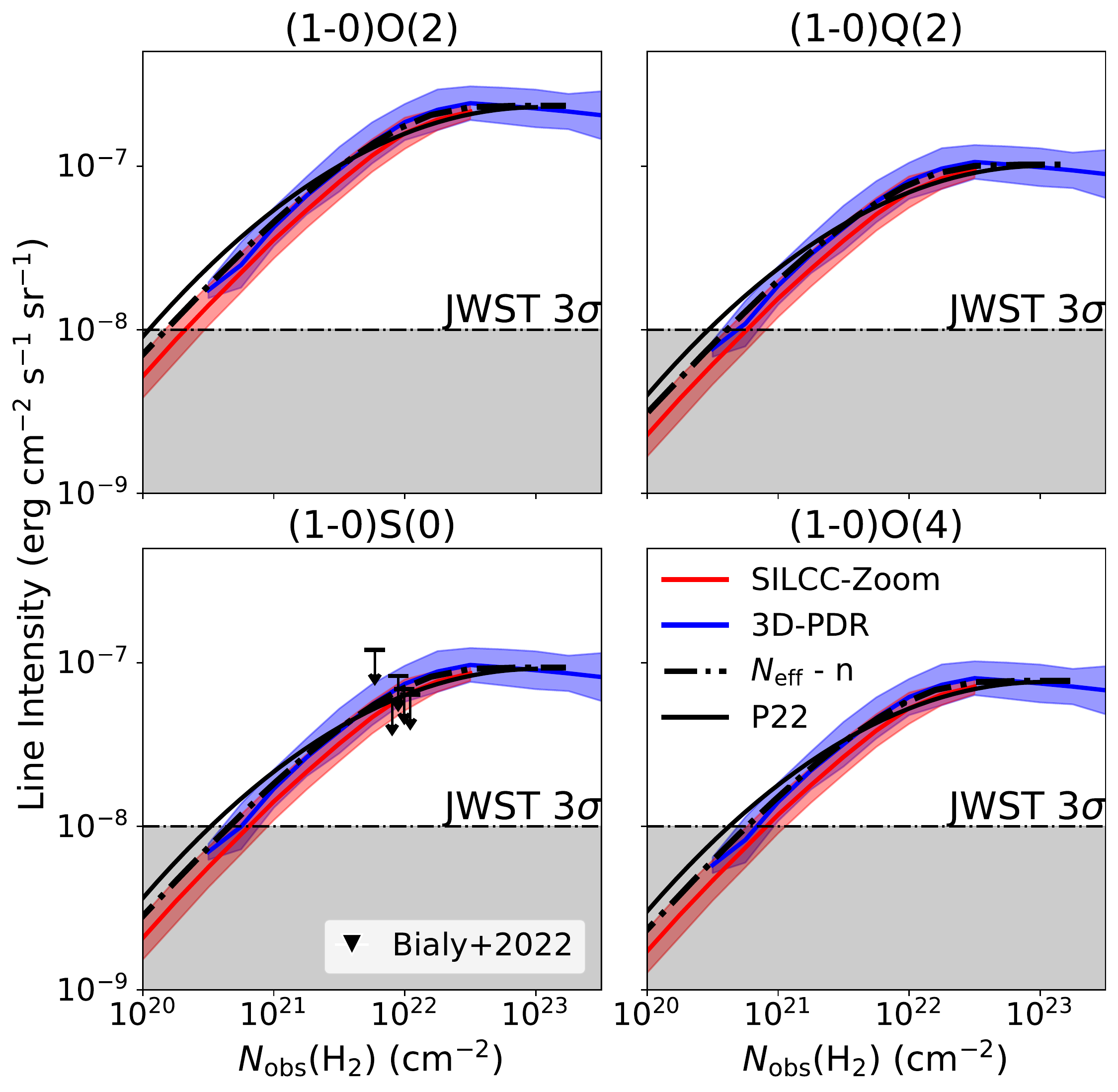}
    \caption{\label{fig:flxDistH2MHD}Same as Fig. \ref{fig:flxDistHMHD} but plotted against the line-of-sight \ce{H2} column density, $N_{\rm obs}(\ce{H2})$. Black solid line corresponds to the calculation in \citetalias{Padovani2022}. Black triangles show the observational upper limits on the (1-0)S(0) line from \citet{Bialy2022}. }
\end{figure}

The impact of different \ce{H2} abundance distributions can be factored out by comparing the \ce{H2} line intensity versus the \ce{H2} column density. Figure \ref{fig:flxDistH2MHD} shows the logarithmic column density bin-averaged line intensity as a function of $N_{\rm obs}(\ce{H2})$ to investigate whether the differences in chemistry evolution, and thus the abundance profiles, are a dominant factor. We find that now the agreement between the Models 3, 4, 5 is strong. We also compare our results to the \citetalias{Padovani2022} model (solid curve) which assumes a constant density, purely molecular slab. Despite the different treatments of the chemistry and density distributions in the various models, there is a good agreement on the line intensity as a function of $N_{\rm obs}(\ce{H2})$. We also show the observational upper limits from \citetalias[]{Bialy2020}, which are consistent with the various models presented.

\section{Conclusions}

The results presented in this paper demonstrate that the assumptions regarding geometry, chemical evolution and density distribution do not play a significant role when the \ce{H2} column density is used along with the \ce{H2} emission lines for constraining the CRIR. However, if the \ce{H2} column density is not constrained, and the total hydrogen column density $N_{\rm obs}$ is used instead, then the various models diverge in their prediction of the \ce{H2} line intensity, especially at low columns, $N_{\rm obs} < 10^{22}$ cm$^{-2}$. This is because the various models assume different density distributions and chemical evolution which result in different H-H$_2$ abundances.

There are a few chemical effects not included in this study which will be investigated in future work. First, we focused primarily here on the role of CRs. However, X-rays can also excite the \ce{H2} lines through secondary electrons produced by X-ray ionizations, in a similar manner as CRs. Second, the \ce{H2} excitation rate and emission assume the \ce{H2} is primarily in the para-\ce{H2} state. However, at low column densities, particularly in regions with enhanced radiation, this assumption may begin to break down \citep[][]{flower2006}. As a result, at low column densities, there may be variations in the \ce{H2} line depending on the ortho-to-para ratio and whether or not state-specific chemistry is included (see Appendix \ref{app:oneD} for a model using an approximate treatment of the ortho-to-para ratio). Third, we have neglected additional H$_2$ excitation processes, such as FUV photo-excitation at the cloud envelopes, and H$_2$ formation pumping. As shown in \citetalias{Bialy2020} (see their Fig. 1, and Eqs.~10, 11) this assumption is valid for clouds sufficiently high in CRIRs or low in FUV irradiation. A comprehensive 3D model that includes these additional H$_2$ excitation mechanisms will be presented in a future work.

To summarize, we have presented synthetic CR induced \ce{H2} line emission maps of four key emission lines (Table \ref{tab:h2params}) for two simulated three-dimensional molecular clouds and several one-dimensional models. These lines are of particular importance: \citetalias{Bialy2020} and \citetalias{Padovani2022} predicted that they trace the CRIR in dense gas and can be simultaneously observed using NIRSPEC on the JWST. We find that the \ce{H2} line intensity as a function of the \ce{H2} column density is relatively insensitive to the assumed density distribution or chemical model. Due to this insensitivity, we recommend the use of the \ce{H2} lines in Table \ref{tab:h2params} for constraining the CRIR in dense gas, in particular using the JWST.

\begin{acknowledgements}
BALG, DS, and SW acknowledge support by the Deutsche Forschungsgemeinschaft (DFG) via the Collaborative Research Center SFB 956 “Conditions and Impact of Star Formation”. TGB acknowledges support from Deutsche Forschungsgemeinschaft (DFG) grant No. 424563772.
SB acknowledges support from the Center for Theory and Computations (CTC) at the University of Maryland. The following {\sc Python} packages were utilized: {\sc NumPy} \citep{numpy}, {\sc SciPy} \citep{scipy}, {\sc Matplotlib} \citep{matplotlib}, {\sc cmocean}. This research has made use of NASA's Astrophysics Data System.
\end{acknowledgements}

\bibliographystyle{aa}
\bibliography{lib} 

\begin{appendix}
\section{One-dimension astrochemical models}\label{app:oneD}
We use three different density distributions: constant $n = 10, 10^3$ cm$^{-3}$ and one following the $N_{\rm eff}$ -- $n$ relation of Eq.~(\ref{eq:avn}). Cosmic-ray attenuation is included following the prescription given in \citet{Gaches2022}, where the hydrogen nuclei column density, $N_{\rm eff}$ is equated with the attenuating column and the ionization rate follows the polynomial fit of the ``$\mathcal{H}$ model'' from \citet{padovani2018}, as described above. 
\begin{figure}[b!]
    \centering
    \includegraphics[width=0.45\textwidth]{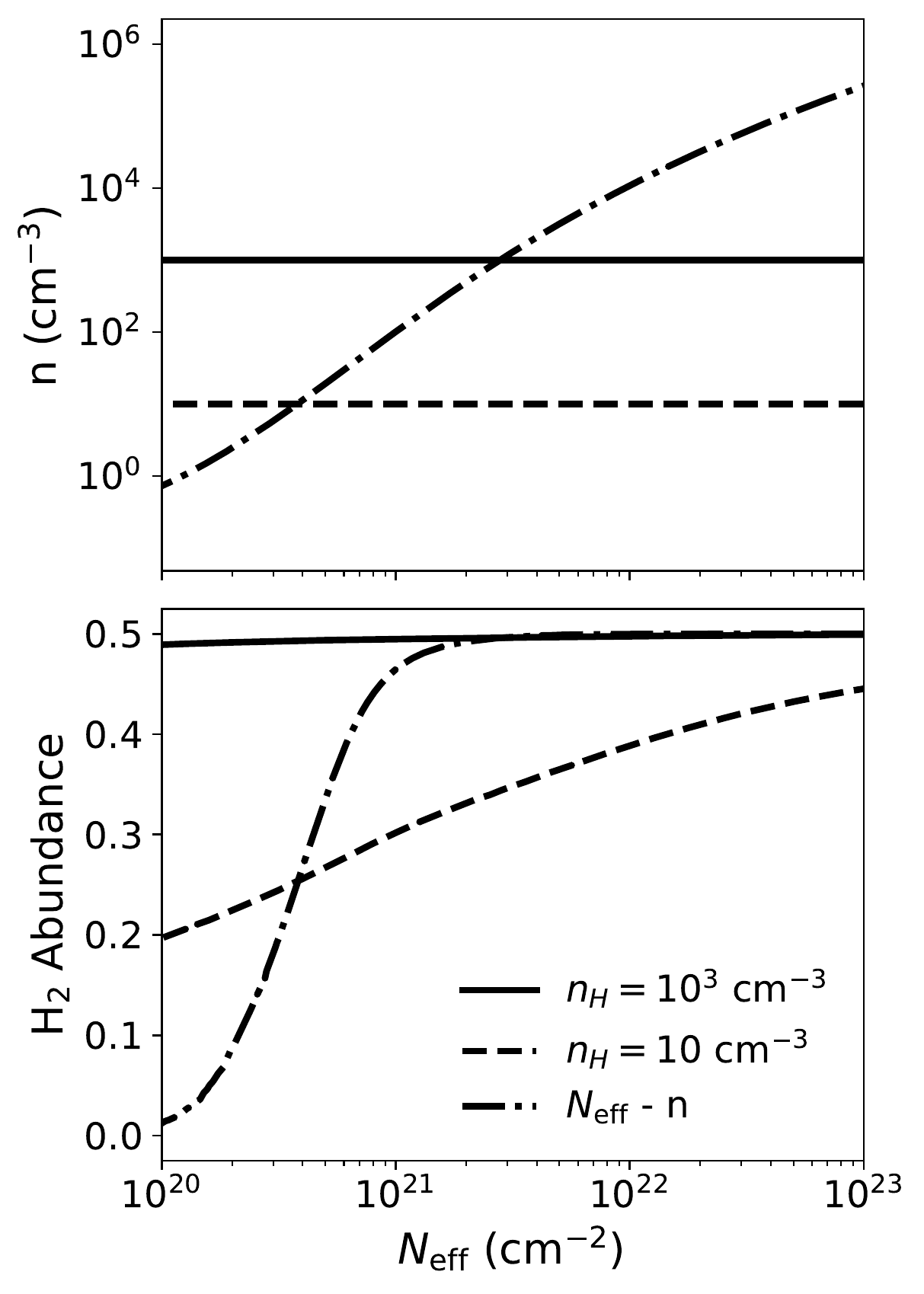}
    \caption{\label{fig:1D_h2_abunds}Total gas density (top) and \ce{H2} abundance (bottom) versus total hydrogen nuclei column density, $N_{\rm eff}$ for three different one-dimensional {\sc 3d-pdr} models.}
\end{figure}

Figure \ref{fig:1D_h2_abunds} shows the hydrogen nuclei number density and \ce{H2} abundance, $x(\ce{H2})\equiv n(\ce{H2})/n$, as a function of hydrogen nuclei column density, $N_{\rm eff}$. The models exhibit a transition from predominantly atomic H to molecular H$_2$, gas
with increasing column density, as the photodissociating FUV radiation is absorbed in the H$_2$ lines and in the dust (for the $n=10^3$ cm$^{-3}$ this transition occurs at small columns, beyond the x-axis lower limit), \citep[][]{sternberg2014, bialy2016, bialy2017}.

\begin{figure}[ht!]
    \centering
    \includegraphics[width=0.45\textwidth]{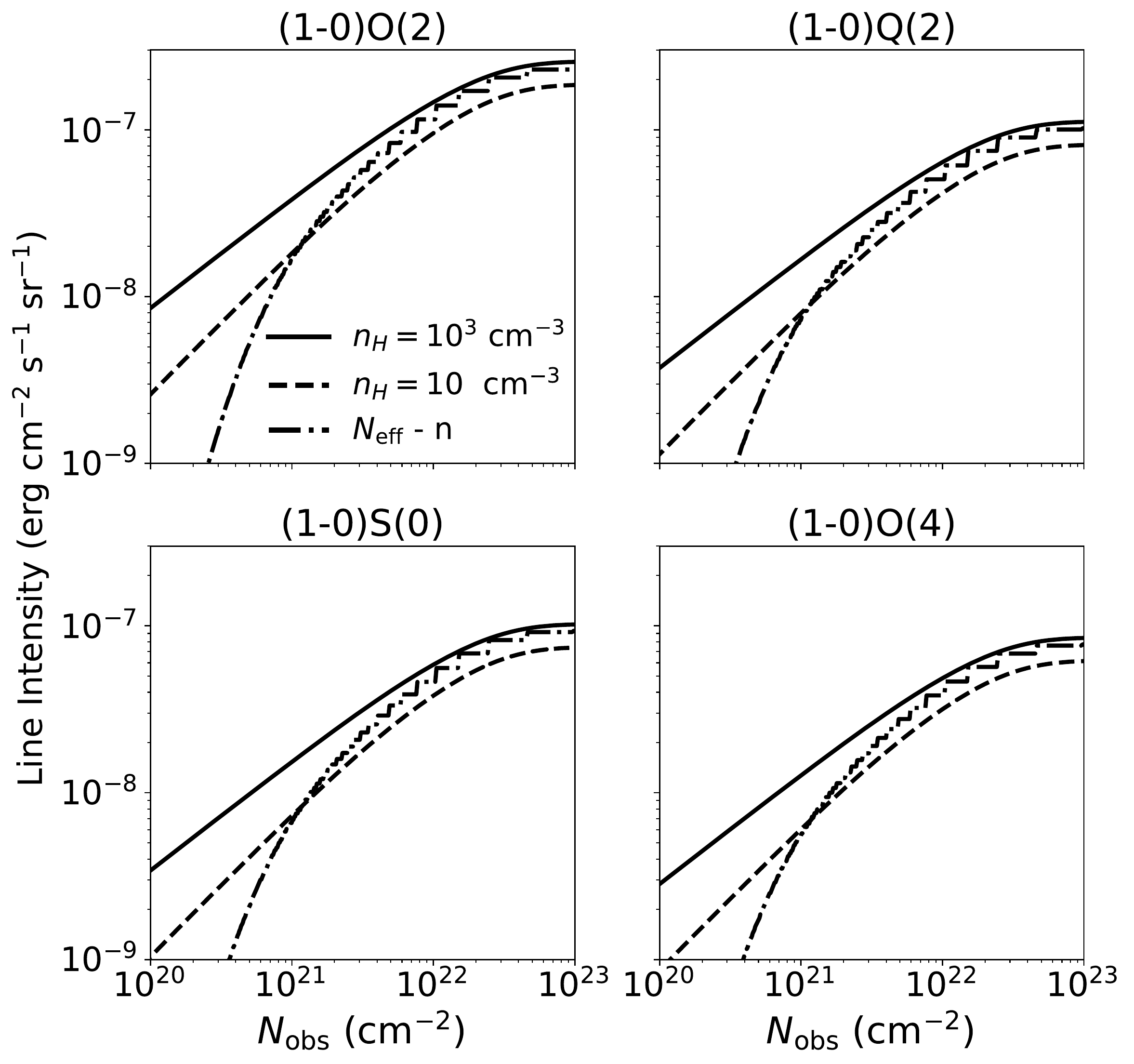}
    \caption{\label{fig:1D_flx}Same as Figure \ref{fig:1d_models_h2} but plotted against the total observed column density, $N_{\rm obs}$.}
\end{figure}

\begin{figure}
    \centering
    \includegraphics[width=0.5\textwidth]{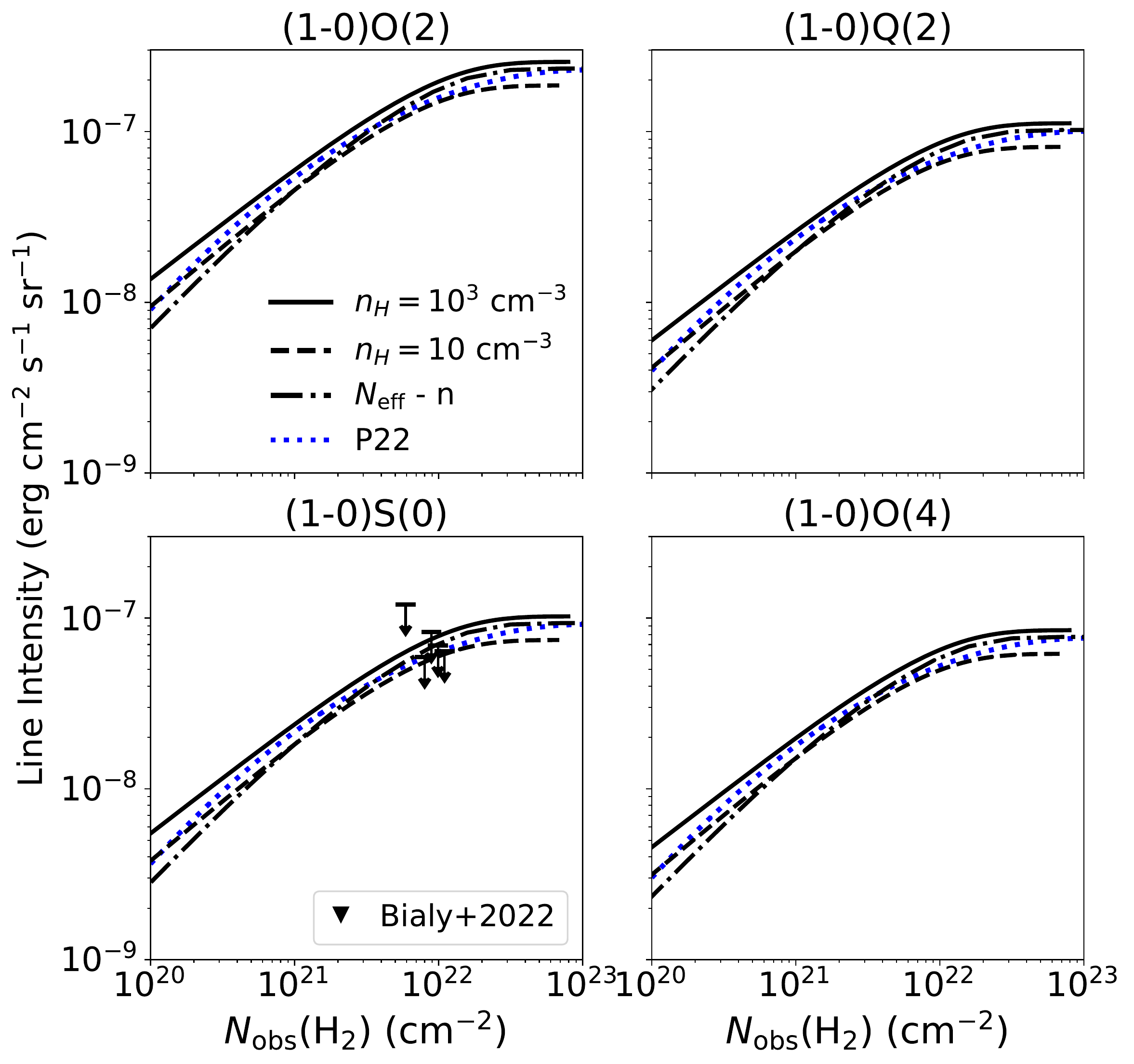}
    \caption{\label{fig:1d_models_h2}Line intensity as a function of \ce{H2} column density for the four \ce{H2} lines considered in this work. Blue dotted line corresponds to the calculation in \citetalias{Padovani2022}. Black triangles show the upper limits on the (1-0)S(0) line from \citet{Bialy2022}. }
\end{figure}

Fig.~\ref{fig:1D_flx} shows that as a function of the total hydrogen column density, the line intensity show significant variation between the models. Figure \ref{fig:1d_models_h2} shows the line flux as a function of the \ce{H2} column density for the one dimensional astrochemical models and exhibits far less variation. The constant density model with $n = 10^3$ cm$^{-3}$ is fully molecular and best matches the \citetalias{Padovani2022} predictions which assumed a fixed \ce{H2} abundance, $x(\ce{H2}) = 0.5$. The $N_{\rm eff}$ -- $n$ model line intensities are dimmer than the $n = 10$ cm$^{-3}$ model at low column densities due to the lower \ce{H2} abundances in this limit. As functions of the \ce{H2} column density, all one dimensional models are in agreement to within an order of magnitude, and consistent with the \citet{Bialy2022} upper limits. 

In our fiducial models, we assume a metallicity of $Z = 1.0 Z_{\odot}$ and that the \ce{H2} is entirely in the para-\ce{H2} spin state. We ran an additional set of four models using the $N_{\rm eff}$ -- $n$ density distribution. Three models use different metallicities: $Z = 0.1 Z_{\odot}$, $Z = 0.5 Z_{\odot}$, and $Z = 2.0 Z_{\odot}$ and the fourth uses $Z = 1.0 Z_{\odot}$ but assumes the \ce{H2} ortho-to-para ratio (OPR) is in thermal equilibrium \citep{flower2006},
\begin{equation}
    {\rm OPR} = 9 \exp(-170.5/T_g),
\end{equation}
where $T_g$ is the gas temperature computed by {\sc 3d-pdr} and was computed in post-processing for the \ce{H2} emission since {\sc 3d-pdr} does not include spin chemistry. This relationship deviates from the expected asymptotic OPR ratio of 3 at high temperatures, but the gas we consider is generally cool (T < 100 K), where this relationship still produces an OPR less than 3. However, the use of this approximation will provide a first indication of the importance of the OPR in determining these line intensities. The \ce{H2} line emissivity is then modified as
\begin{equation}
    \epsilon_{ul} = \alpha_{ul} \frac{E_{ul}}{4\pi} \zeta_{\rm exc,u} \left ( \frac{n(\ce{H2})}{1 + {\rm OPR}} \right ) \, .
\end{equation}
The dust opacity and \ce{H2} formation rates are linearly scaled with metallicity. Further, the metallicity impacts the heating and cooling due to the changes in the abundances of metals and dust.

\begin{figure}[htb!]
    \centering
    \includegraphics[width=0.45\textwidth]{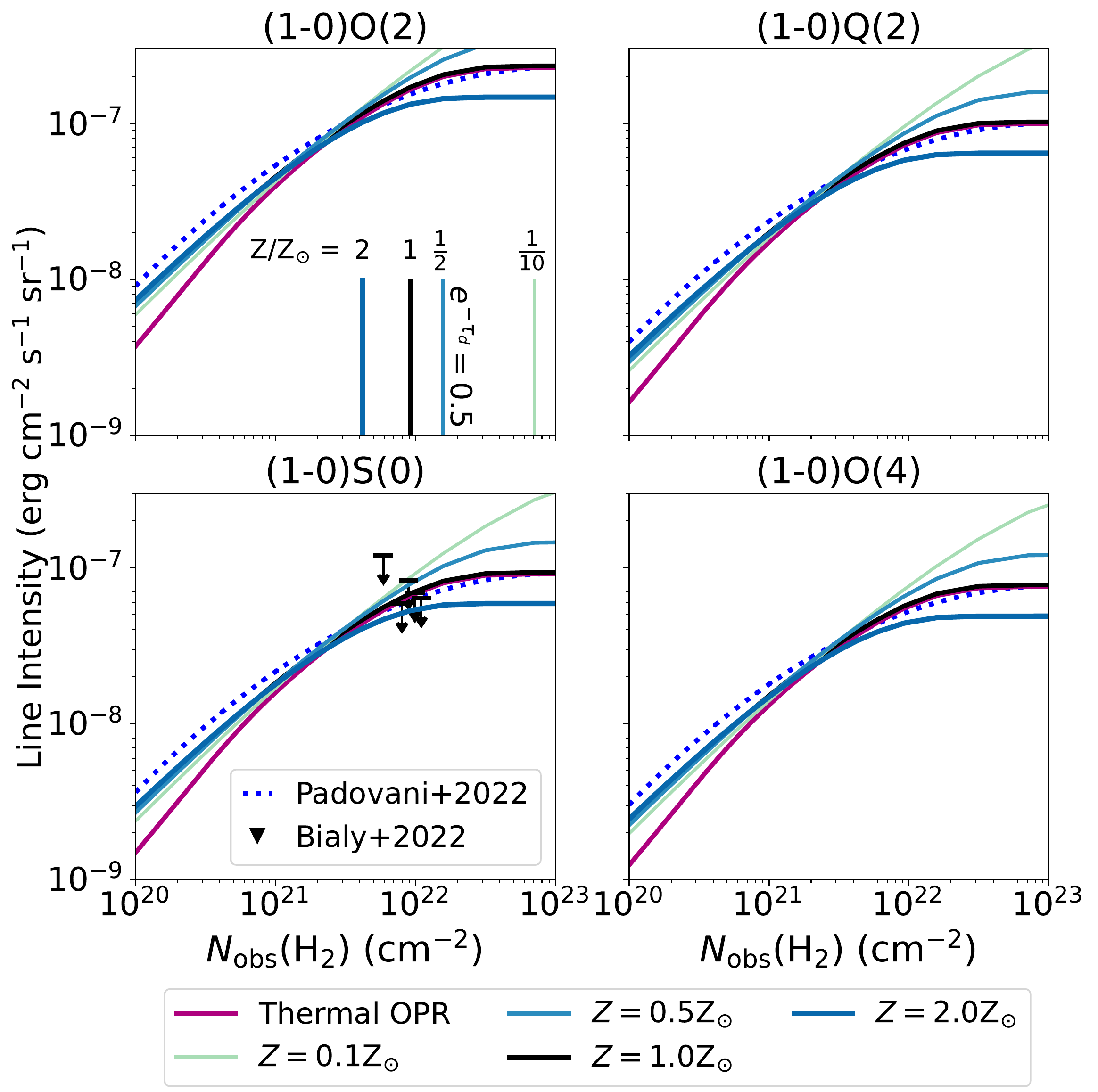}
    \caption{\label{fig:1D_flx_var}Same as Figure \ref{fig:1d_models_h2} but plotting the $N_{\rm eff}$ -- $n$ model for different metallicities and a model using an ortho-to-para ratio in thermal equilibrium. The vertical lines in the top-left plot denote the \ce{H2} column density where $\rm{e}^{-\tau_d} = 0.5$.}
\end{figure}

Figure \ref{fig:1D_flx_var} shows the \ce{H2} line intensities versus \ce{H2} column density for these four different models along with the fiducial $N_{\rm eff}$ -- $n$ model. At low \ce{H2} column densities, there is little deviation, although the ``Thermal OPR'' model shows a slight decrease in line intensity at low column densities (and thus higher temperatures) due to \ce{H2} also being in ortho-\ce{H2} spin state. At high column densities, the line intensities asymptote to different values due to lower (higher) metallicities have their $\tau_d = 1$ surface deeper (shallower) in the clouds.

\end{appendix}

\end{document}